\documentclass[aps,amssymb,eqsecnum,preprint,12pt]{revtex4}
\usepackage{bm}
\usepackage{graphicx,color}
\usepackage{latexsym}
\usepackage{bm,amsmath}
\begin{document}

\newcommand{\M}{\cal M}
\newcommand{\N}{\cal N}
\newcommand{\C}{\cal C}
\def\bra#1{\mathinner{\langle{#1}|}}
\def\ket#1{\mathinner{|{#1}\rangle}}
\def\braket#1{\mathinner{\langle{#1}\rangle}}

\vspace*{1cm}
\title{
Stationary Rotating Strings as Relativistic Particle Mechanics
\vspace{1cm}
}

\author{
Kouji Ogawa\footnote{Present address: 
Observations Division,  
Fukuoka District Meteorological Observatory, Fukuoka}, 
 Hideki Ishihara\footnote{
E-mail:ishihara@sci.osaka-cu.ac.jp}, 
 Hiroshi Kozaki$^{1}$
 Hiroyuki Nakano$^2$, and
 Shinya Saito\footnote{Present address:
Observations Division, Meteorological Instruments Center, Ibaraki}
\bigskip
}
%\address
\affiliation{
Department of Mathematics and Physics, Graduate School of Science, 
Osaka City University, Osaka 558-8585, Japan
\bigskip
\\
{}$^1$Department of Applied Chemistry and Biotechnology, 
  Niigata Institute of Technology, 
  Kashiwazaki, Niigata 945--1195 Japan
\bigskip
\\ 
{}$^2$Center for Computational Relativity and Gravitation, 
School of Mathematical Sciences, 
Rochester Institute of Technology, 
Rochester, New York 14623, USA
\vspace{2cm}
}

\preprint{OCU-PHYS 295}
\preprint{AP-GR 57}
\date{\today\vspace{2cm}
}

\begin{abstract}
Stationary rotating strings can be viewed as geodesic motions in 
appropriate metrics on a two-dimensional space. 
We obtain all solutions describing stationary rotating strings in 
flat spacetime as an application. 
These rotating strings have infinite length with various wiggly shapes. 
Averaged value of the string energy, the angular momentum and the 
linear momentum along the string are discussed. 
\end{abstract}

\maketitle

%%%%%%%%%%%%%%%%%%%%%%%%%%%%%%%%%%%%
\section{Introduction}
%%%%%%%%%%%%%%%%%%%%%%%%%%%%%%%%%%%%

Cosmic strings are topological defects produced by U(1) symmetry breaking 
in the unified field theories, which is believed to occur 
in the early stage of the universe\cite{Kibble76}. 
(See also \cite{VandS} for a review.) 
Verification of the existence of cosmic strings 
is a strong evidence of the occurrence of vacuum phase transition 
in the universe. 
For detection of the cosmic strings, clarification of the string motion 
is an important task. 

For the cosmic strings in the framework of gauge theories, 
the reconnection probability is essentially one\cite{Shellard}. 
When strings cross, they reconnect and reduce their total length. 
The closed loops are produced by self-intersections of long strings, 
and loops decay through gravitational radiation. It is known that 
the strings evolve in a scale-invariant manner
\cite{Albrecht,Bennett,Allen}. 
There are renewed interest of cosmic strings, recently, in 
relation to the spacetime geometry of the compact extra dimensions 
of fundamental string theories including 
branes\cite{Sarangi:2002yt, Dvali:2003zj, Copeland:2003bj}. 
A detailed investigation\cite{Jackson:2004zg}  suggests that 
the reconnection probability of this type of strings is 
considerably suppressed. In this case, the strings in the 
universe are practically stable. 
Then, it is a natural question that what is the final state of 
cosmic strings? 

It is well known that the final state of black holes is the Kerr spacetime, 
which is a stationary state. 
We guess analogously that the final state of cosmic strings would be 
a stationary string. 
The stationary string is defined as the world surface which is 
tangent to a time-like Killing vector field 
if one neglects the thickness and the gravitational effects of a string. 
There are many works on stationary strings in various stationary 
target spacetimes\cite{BurdenSRS, VLS, FSZH, Anderson}.

We consider infinitely long and stationary rotating strings in the Minkowski 
spacetime whose dynamics is governed by the Nambu-Goto action in this paper. 
Though this subject is already studied by Burden and Tassie\cite{BurdenSRS} 
in a different motivation, and 
also investigated in the literatures\cite{VLS, FSZH, Anderson},
it is worth clarifying physical properties of stationary 
rotating strings completely because 
the stationary rotating strings have a rich variety of configuration 
even in Minkowski spacetime.

A stationary string is an example of cohomogeneity-one string. 
The world surface of cohomogeneity-one string is 
tangent to a Killing vector field of a target space, 
equivalently, the world surface is foliated by the orbits of 
one-parameter group of isometry. 
All possible cohomogeneity-one strings in Minkowski spacetime 
are classified into seven families\cite{IshiharaKozaki}. 
Stationary rotating strings in Minkowski spacetime belong to one 
family of them. 
Solving equations of motion for cohomogeneity-one string is reduced to 
finding geodesic curves in a three-dimensional space with the metric 
which is determined by the 
Killing vector\cite{FSZH, IshiharaKozaki}. 

We would find an analogy between the system of cohomogeneity-one 
string and the system of Bianchi cosmologies. 
In the Bianchi cosmologies, universe is foliated by homogeneous 
time slices, i.e., spacetime is cohomogeneity-one. 
The dynamics of Bianchi cosmologies is 
regarded as the one of a relativistic particle. 
Similarly, the dynamics of each family of cohomogeneity-one 
string can be identified by the one of a particle moving in a curved 
space specified by the geometrical symmetry of 
the string\cite{IshiharaKozaki}. 
We perform this procedure of identification for stationary rotating strings 
as the first step, in this paper. 
As a result, we show that the system of stationary rotating 
strings can be formulated as the dynamical system of particles moving 
along geodesics in two-dimensional curved spaces. 
It is important to clarify the geometrical structure of the two-dimensional 
space to understand the stationary rotating strings. This viewpoint 
is another motivation of this paper.

This paper is organized as follows. 
In Section 2 we formulate the system of stationary rotating strings 
in Minkowski spacetime as dynamical systems of particles. 
In Section 3 general solutions for the system are 
presented. In Section 4 we examine physical properties of the 
stationary rotating strings. Finally, Section 5 is devoted to 
discussions.

%%%%%%%%%%%%%%%%%%%%%%%%%%%%%%%%%%%%%%%%%%%%%
\section{Stationary Rotating Strings in Minkowski Spacetime}
\label{2_string}
%%%%%%%%%%%%%%%%%%%%%%%%%%%%%%%%%%%%%%%%%%%%%

%%%%%%%%%%%%%%
\subsection{Equations of Motion for Cohomogeneity-One Strings}
%%%%%%%%%%%%%%

A string is a two-dimensional world surface $\Sigma$ in 
a target spacetime $\M$. 
The embedding of $\Sigma$ in $\M$ is described by
$$
 x^\mu = x^\mu (\zeta^a), 
$$ 
where $ x^\mu $ are coordinates of $\M$ 
and $\zeta^a ~(\zeta^0=\tau ,\zeta^1=\sigma)$ are two parameters on $\Sigma$. 
We assume that the dynamics of string is governed by the Nambu-Goto action 
\begin{align}
	S = -\mu \int_\Sigma d^2 \zeta \ \sqrt{- \gamma }, 
\label{NG} 
\end{align}
where $\mu$ is the string tension and 
$\gamma $ is the determinant of the induced metric 
on $\Sigma$ given by
\begin{align}
	\gamma_{ab} 
		= g_{\mu\nu}{\partial_a}x^\mu 
			{\partial_b}x^\nu,
\end{align}
where $g_{\mu\nu}$ is the metric of $\M$. 

Let us consider that the metric $g_{\mu\nu}$ of $\M$ 
possesses isometries generated by Killing vector fields. 
If $\Sigma$ is tangent to one of the Killing vector fields of $\M$, 
say $\xi$, 
we call the world surface $\Sigma$ a cohomogeneity-one string associated 
with the Killing vector $\xi$. 
The stationary string is one of the example 
of the cohomogeneity-one string, where the Killing vector is timelike. 

The  group action of isometry generated by $\xi$ 
on $\M$ defines the orbit of $\xi$. 
The metric on the orbit space is naturally 
introduced by the projection tensor with respect to $\xi$, 
\begin{equation}\begin{aligned}
	h_{\mu\nu}:=g_{\mu\nu}-\xi_\mu\xi_\nu/f, 
\label{hab}
\end{aligned}\end{equation}
where $f:=\xi^\mu\xi_\mu$.  
The action for the cohomogeneity-one string 
associated with a Killing vector $\xi$ is reduced to 
the action for a curve $\C$ 
in the form\cite{FSZH, IshiharaKozaki} 
\begin{equation}\begin{aligned}
	S=\int_{\C} \sqrt{-f h_{\mu\nu}dx^\mu dx^\nu}. 
\label{line_action}
\end{aligned}\end{equation}
The metric $h_{\mu\nu}$ has Euclidean signature in the case of timelike 
Killing vector, $f<0$,  
and Lorentzian signature in the case of spacelike Killing vector, $f>0$. 

The action \eqref{line_action} gives the length of $\C$ 
with respect to the metric $-f h_{\mu\nu}$ on the orbit space of $\xi$. 
Therefore, the problem for finding solutions of cohomogeneity-one string 
associated with $\xi$ reduces to 
the problem for solving three-dimensional geodesic equations 
with respect to the metric $-f h_{\mu\nu}$.

%%%%%%%%%%%%%%%%%%%%%%%%%%%%%%%%%%%%%%%%%%%%%%%%
\subsection{Equations of Motion for Stationary Rotating Strings }
%%%%%%%%%%%%%%%%%%%%%%%%%%%%%%%%%%%%%%%%%%%%%%%%

We consider stationary rotating strings in Minkowski spacetime 
with the metric in the cylindrical coordinate 
for an inertial reference frame, 
\begin{align}
 ds^2 = -d\bar t^2 + d\bar\rho^2 + \bar \rho^2 d\bar \varphi^2 +d\bar z^2. 
\label{MinMetCyl} 
\end{align}
The world surface is tangent to the Killing vector field
\begin{equation}\begin{aligned}
	\xi = \partial_{\bar t} +\Omega \partial_{\bar\varphi}, 
\label{Killing}
\end{aligned}\end{equation}
where $\Omega$ is a constant denoting the angular velocity. 
We introduce a new coordinate
\begin{equation}\begin{aligned}
	x^\mu 
		=( {t},  {\rho}, {\varphi},  {z})
		=(\bar t, \bar \rho, \bar \varphi-\Omega \bar t, \bar z)
\end{aligned}\end{equation}
such that 
\begin{equation}\begin{aligned}
	\xi = \partial_{ {t}} . 
\end{aligned}\end{equation}
The coordinate system $ {x}^\mu$ gives the rigidly rotating 
reference frame with the angular velocity $\Omega$. 
In this coordinate system, we rewrite the metric of Minkowski spacetime as
\begin{align}
 ds^2 &=  {g}_{\mu \nu} d {x}^\mu d {x}^\nu  \cr
	&= -(1-\Omega^2  {\rho}^2)\left( d {t}-\frac{\Omega {\rho}^2 }
			{1-\Omega^2  {\rho }^2 }d {\varphi}\right)^2 
			+ d {\rho}^2 
		+ \frac{ {\rho}^2}{1-\Omega^2  \rho^2}d {\varphi}^2 
			+ d {z}^2. 
\label{tildeg}
\end{align}
The norm of the Killing vector is given by 
$f=\xi_\mu \xi^\mu=-(1-\Omega^2  \rho^2)$, and 
the metric of orbit space defined by \eqref{hab} is 
\begin{align}
 ds_{(3)}^2 = h_{ij} dx^i dx^j 
		= d \rho^2 
		+ \frac{\rho^2}{1-\Omega^2 \rho^2} d \varphi^2 
		+dz^2, \quad (i,j=1,2,3).
\label{met3}
\end{align}
The singular point $\rho=1/|\Omega|$ of the three-dimensional 
metric \eqref{met3}
is the light cylinder where the rotation velocity becomes the 
light velocity. 
For a stationary rotating string the Killing vector $\xi$ is 
timelike on $\Sigma$, 
then the string should stay in the region $ \rho < 1/|\Omega|$. 
The Nambu-Goto equation for the stationary rotating string 
whose world surface is tangent to $\xi$ in Minkowski spacetime 
is reduced to the three-dimensional geodesic equation 
in the metric 
\begin{align}
 d {\tilde s}_{(3)}^2 = \tilde h_{ij} d {x}^i d {x}^j 
				= -f h_{ij} d {x}^i d {x}^j 
		= (1-\Omega^2  {\rho }^2 ) d {\rho}^2 
		+  {\rho}^2 d {\varphi}^2 
		+ (1-\Omega^2  {\rho }^2 ) d {z}^2 . 
\label{eff3metric} 
\end{align}
The world surface of the string is spanned 
by the geodesic curve, say $ {x}^k (\sigma)$, 
and the Killing vector $\xi$. 
Then, by using a parameter choice 
\begin{equation}\begin{aligned}
	\tau = t, 
\end{aligned}\end{equation}
we can give the embedding of $\Sigma$, 
$ {x}^\mu(\tau, \sigma)$, in the form
\begin{align}
	 {x}^\mu (\tau , \sigma ) = 
 		\left\{\begin{array}{l}
			 {t} (\tau ) = \tau \\
			 {x}^k (\sigma ) 
				= ( \rho(\sigma),  \varphi(\sigma),  {z}(\sigma)). 
 		\end{array}
 \right. 
 \label{ParAnsSR}
\end{align}

%%%%%%%%%%%%%%%%%%%%%%%%%%%%%%%%%%%%%%%%%%%%%%%%%%%%%%%%%%%%%%%%%%%%%
\subsection{Reduced system}
%%%%%%%%%%%%%%%%%%%%%%%%%%%%%%%%%%%%%%%%%%%%%%%%%%%%%%%%%%%%%%%%%%%%%

We should obtain the geodesics in the three-dimensional orbit space. 
The action \eqref{line_action} for the geodesics with respect to 
the metric \eqref{eff3metric} is equivalent to 
the action 
\begin{align}
 S 	&=\frac12 \int \left[ \frac{1}{N} \tilde h_{ij}  x'^i x'^j 
		+N\right]d\sigma 
\label{P_action_formal}
\\
  &= \frac{1}{2}\int \left[\frac{1}{N}\left\{(1-\Omega^2{\rho }^2 ){\rho}'^2 
		+  {\rho}^2  {\varphi}'^2 
		+ (1-\Omega^2{\rho }^2){z}'^2\right\} +N\right]d\sigma, 
\label{P_action}
\end{align}
where the prime denotes the derivative with respect to $\sigma$, 
and $N$ is an arbitrary 
function of $\sigma$ which is related to the reparametrization 
invariance of the curve. 

%%%%%%%%%%%%%%%%%%%%%%%%%%%%%%%%%

In the action \eqref{P_action}, $z$ and $\varphi$ are 
cyclic (ignorable) coordinates.   
Here, using the fact that the momentum conjugate to $z$ is conserved 
we pay attention to curves which is projected on $\rho-\varphi$  plane. 
We can interpret the projected curves in two ways; One is geodesics in a 
two-dimensional curved space, and the other is spatial orbits of a particle 
moving in a potential in two-dimensional flat space. 
We proceed along the first viewpoint, and also give brief discussion 
with the second view in Appendix A. 

%---------------------------------------------------------

Action \eqref{P_action_formal} is written in the Hamilton form 
\begin{equation}\begin{aligned}
	S = \int \left[p_i {  x}'^i 
			- N{\cal H} \right] d\sigma ,
\label{H_action}
\end{aligned}\end{equation}
where the canonical conjugate momentum $p_i$ with respect to $  x^i$ is 
\begin{equation}\begin{aligned}
		p_i	= \frac{1}{N}\tilde h_{ij}   x'^j.
\label{momentum}
\end{aligned}\end{equation}
Variation of \eqref{H_action} by $N$ leads that the Hamiltonian $N{\cal H}$ 
is vanishing, i.e.,
\begin{equation}\begin{aligned}
	{\cal H} = \frac{1}{2}\left( \tilde h^{ij} p_i p_j -1 \right)=0.
\label{Hconstraint}
\end{aligned}\end{equation}

Since the momentum $p_z$, which is conjugate to the cyclic coordinate $z$, 
takes a constant value, say $q$. 
The action \eqref{H_action} reduces to
\begin{equation}\begin{aligned}
	S = S_0+ q   z - \int N{\cal H}  d\sigma , 
\end{aligned}\end{equation}
where
\begin{equation}\begin{aligned}
	S_0 = \int p_A x'^A  d\sigma, \qquad (x^A= \rho,  \varphi). 
\label{Sred}
\end{aligned}\end{equation} 
When we concentrate on the curves that satisfy $p_z=q$ and ${\cal H}=0$, 
according to Maupertuis' principle, the projected curve is the curve  
that extremizes the abbreviated action, $S_0$. 

With the help of \eqref{momentum} the Hamiltonian constraint is written in the form 
\begin{align}
			0={\cal H} &=\frac{1}{2}\left( \tilde h^{AB}p_A p_B 
		+ \tilde h^{zz}q^2 -1 \right) \cr
				&=\frac{1}{2}\left(\frac{1}{N^2}\tilde h_{AB}  x'^A   x'^B 
				+ \tilde h^{zz}q^2 -1 \right). 
\end{align}
Then, we have
\begin{equation}\begin{aligned}
 	 \tilde h_{AB}dx^A dx^B + (\tilde h^{zz}q^2 -1) N^2 d\sigma^2 =0.
\label{zeta}
\end{aligned}\end{equation}
Inserting \eqref{momentum} and \eqref{zeta} into 
\eqref{Sred}, we obtain Jacobi's form of the abbreviated action, 
\begin{align}
	S_0 &=\int \frac{1}{N}\tilde h_{AB} x'^A x'^B d\sigma \cr
	    &= \int \sqrt{(1-\tilde h^{zz}q^2)\tilde h_{AB}dx^A dx^B}.
\end{align}
Therefore, the action \eqref{P_action} reduces to the geodesic action  
\begin{equation}\begin{aligned}
	S_0 = \int \sqrt{h^{red}_{AB}dx^A dx^B}, 
\label{2dim-action}
\end{aligned}\end{equation}
where the metric of reduced two-dimensional space is given as 
\begin{equation}\begin{aligned}
	h^{red}_{AB}= (1- \tilde h^{zz}q^2) \tilde h_{AB}.
\label{hred}
\end{aligned}\end{equation}

By variation of $S$ by $q$, we have 
\begin{equation}\begin{aligned}
	  z = -\frac{\partial S_0}{\partial q}.
\label{z}
\end{aligned}\end{equation}
The geodesic curves extremizing \eqref{2dim-action} together with \eqref{z} 
determine the full three-dimensional geodesics in the orbit space.

For the stationary rotating strings, 
the metric of two-dimensional curved space on which 
we seek geodesics is expressed explicitly in the form
\begin{align}
	ds_{red}^2 &= h^{red}_{AB} dx^A dx^B 
		= (1- \tilde h^{zz}q^2) \tilde h_{AB} dx^A dx^B  \cr
		&= (1-q^2-\Omega^2 \rho^2)
		\left(d \rho^2 +\frac{ \rho^2}{1-\Omega^2 \rho^2}d  \varphi^2\right).
\label{2dim-metric}
\end{align}
The action for geodesics \eqref{2dim-action} is equivalent to
\begin{align}
	S_0 &= \frac{1}{2}\int \left[ \frac{1}{N_{(2)}} (1-q^2-\Omega^2 \rho^2)
		\left( \rho'^2 +\frac{ \rho^2}{1-\Omega^2 \rho^2}\varphi'^2\right)
		+ N_{(2)} \right]d\sigma, 
\label{2dim-P_action}
\end{align}
where $N_{(2)}$ is an arbitrary function of $\sigma$ 
which is related with the freedom of parametrization on a curve.

By variation of the action \eqref{2dim-P_action} by $N_{(2)}$ and $ \varphi$ 
we have 
\begin{align}
		&(1-q^2-\Omega^2 \rho^2)
		\left(  \rho'^2 
		+\frac{ \rho^2}{1-\Omega^2 \rho^2}
		 \varphi'^2\right) =N_{(2)}^2, 
\label{2dim-norm}\\
		&\frac{(1-q^2-\Omega^2 \rho^2) \rho^2}
				{1-\Omega^2 \rho^2} \varphi'=N_{(2)}\frac{\ell}{|\Omega|}, 
\label{2dim-ell}
\end{align}
where $\ell$ is a constant. 

Eliminating $\varphi'$ from \eqref{2dim-norm} and \eqref{2dim-ell} we have
\begin{align}
	 {\rho}'^2 = \frac{N_{(2)}^2}{1-q^2-\Omega^2 \rho^2} 
		\left[1-\frac{\ell^2(1-\Omega^2 \rho^2) }
		{(1-q^2-\Omega^2 \rho^2)\Omega^2 {\rho}^2} 
		\right].
\label{integral}
\end{align}
We see that $ \rho$ is bounded as 
$\rho_{min}\leq  \rho \leq  \rho_{max}$ because the right hand side of 
\eqref{integral} should be positive, where
\begin{equation}\begin{aligned}
	& \rho_{min}^2 = \frac{1}{2\Omega^2}\left(
			1+\ell^2-q^2 
	- \sqrt{(1+\ell +q)(1+\ell -q)(1-\ell +q)(1-\ell -q)}\right), 
\\
	& \rho_{max}^2 = \frac{1}{2\Omega^2}\left(
			1+\ell^2-q^2
	+ \sqrt{(1+\ell +q)(1+\ell -q)(1-\ell +q)(1-\ell -q)}\right),
\label{minmax}
\end{aligned}\end{equation}
with
\begin{equation}\begin{aligned}
 &1+\ell^2 -q^2 \ge 0, \\
 &(1+\ell +q)(1+\ell -q)(1-\ell +q)(1-\ell -q) \geq 0. 
\end{aligned}\end{equation}
Allowed region of the parameters are shown in Fig.1.

If $q\neq 0$, it is easy to show that
\begin{equation}\begin{aligned}
	(1-\Omega^2 \rho_{min}^2)(1-\Omega^2 \rho_{max}^2) > 0.
\end{aligned}\end{equation}
It means that there are two cases: 
\begin{equation}\begin{aligned}
	\mbox{(i)}&\quad (1-\Omega^2 \rho_{min}^2)>0 \quad 
		\mbox{and} \quad (1-\Omega^2 \rho_{max}^2) > 0 \quad 
&\mbox{for} &\quad 1-\ell^2 +q^2 > 0, \\
	\mbox{(ii)}&\quad (1-\Omega^2 \rho_{min}^2)<0 \quad 
		\mbox{and} \quad (1-\Omega^2 \rho_{max}^2) < 0 \quad 
&\mbox{for} &\quad 1-\ell^2 +q^2 < 0 .
\end{aligned}\end{equation}
In the first case, the Killing vector field $\xi$ is 
timelike everywhere on the world surface $\Sigma$, 
while $\xi$ is spacelike everywhere on $\Sigma$ in the second case. 
In the both cases (i) and (ii), the strings are rigidly rotating\cite{FSZH}. 
The solutions in the case (i) give the stationary rotating strings.

%%%%%%%%%%%%%%%%%%%%%%%%%%%%%%%%%%%%%%%%%%%%%%%%%%%%%%%%%%%%
\begin{figure}
 \includegraphics[width=6cm]{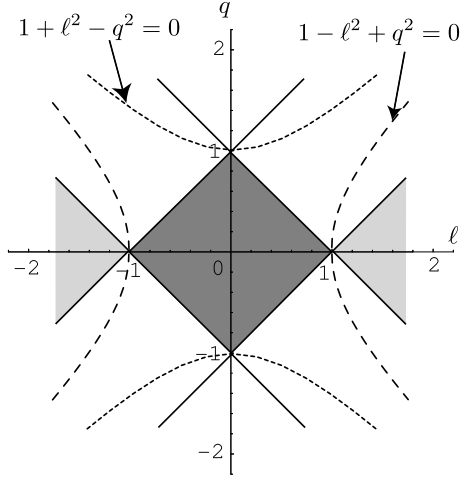}
\caption{Parameter regions. 
The shaded areas are the parameter regions 
for the rigidly rotating strings, the cases (i) and (ii). 
The dark shaded area is the parameter region 
for the stationary rotating strings, the case (i).  
}
\label{lq}
\vspace{5mm}
\end{figure}

%%%%%%%%%%%%%%%%%%%%%%%%%%%%%%%%%%%%%%%%%%%%%%%%%%%%%%%%%%%%

There are three singularities of $h^{red}_{AB}$ given by \eqref{2dim-metric}, 
$\rho=0$, $\rho=\rho_q :=\sqrt{1-q^2}/|\Omega|$ and 
$\rho=1/|\Omega|$.  
For the stationary string, $0 \leq  \rho \leq \rho_{max} \leq \rho_q$. 
The point $ \rho=0$ is a fixed point of the rotational isometry generated by 
the Killing vector $\partial/\partial \varphi$. 
When $q\neq 0$ the point $ \rho=\rho_q$ is also 
a fixed point of the rotation, 
that is, there are two centers in this space. 
When $q=0$ the point $ \rho=\rho_q$ becomes a circle. 

\begin{figure}[ht]
 \includegraphics[width=7cm]{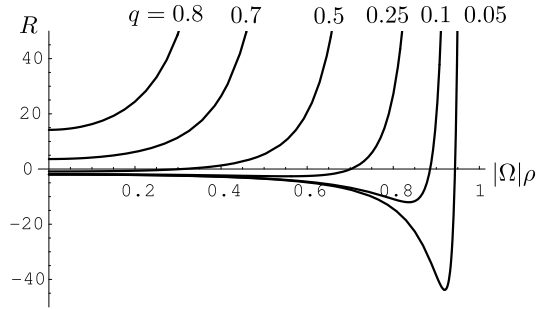}
\caption{Scalar curvature of the reduced two-dimensional space. 
}
\label{curvature}
\end{figure}

The scalar curvature of the metric \eqref{2dim-metric} is calculated as
\begin{equation}\begin{aligned}
	R=\frac{-6 q^4 + 2 \left(\Omega^2 \rho^2+4\right)\left(1-\Omega^2 \rho^2\right)
   q^2-2 \left(1-\Omega^2 \rho^2\right)^3}
		{\left(1-\Omega^2 \rho^2\right)^2 \left(1-q^2-\Omega^2 \rho^2\right)^3}.
\end{aligned}\end{equation}
One of the centers $ \rho=\rho_q$ is the curvature singularity, 
while the other center $ \rho=0$ is regular point. 
The scalar curvature vanishes at $ \rho=0$ when $q^2=1/3$. 
When $q^2>1/3$, the scalar curvature is positive everywhere, while 
the curvature becomes negative near $ \rho=0$ when $q^2<1/3$. 
When $q$ is small enough, the scalar curvature has the minimum point near 
$ \rho=\rho_q$. 
Fig.\ref{curvature} shows the scalar curvature as 
the function of $|\Omega|\rho$. 

%%%%%%%%%%%%%%%%%%%%%%%%%%%%%%%%%%%%%%%%%%%%%%%%%%%%%%%%%%%%
%%%%%%%%%%%%%%%%%%%%%%%%%%%%%%%%%%%%%%%%%%%%%%%%%%%%%%%%%%%%

%%%%%%%%%%%%%%%%%%%%%%%%%%%%%%%%%%%
\section{Solutions for Stationary Rotating Strings}
%%%%%%%%%%%%%%%%%%%%%%%%%%%%%%%%%%%

If we choose the function $N_{(2)}$ as
\begin{equation}\begin{aligned}
	N_{(2)}=1-q^2-\Omega^2\rho^2, 
\label{2dim-N}
\end{aligned}\end{equation}
two first order differential equations \eqref{2dim-ell} and \eqref{integral} 
become
\begin{align}
		&\varphi'=\frac{\ell}{|\Omega|}\frac{1-\Omega^2\rho^2}{\rho^2}, 
\label{eq-phi}\\
	&{\rho}'^2 = (1-q^2-\Omega^2\rho^2) 
		-\frac{\ell^2(1-\Omega^2\rho^2) }
		{\Omega^2{\rho}^2}.
\label{eq-rho}
\end{align}
We also see that \eqref{z} with \eqref{2dim-P_action} 
simply gives
\begin{equation}\begin{aligned}
	 z= q\sigma.
\label{zsigma}
\end{aligned}\end{equation}
The change of signature of parameters: 
\begin{equation}\begin{aligned}
	\ell \to - \ell, \quad q \to - q, \quad\mbox{and}\quad \Omega \to - \Omega,
\end{aligned}\end{equation}
can be absorbed by the inversion of coordinates:
\begin{equation}\begin{aligned}
	\varphi \to - \varphi, \quad z \to - z, \quad\mbox{and}\quad t \to - t,
\end{aligned}\end{equation}
respectively. Then we restrict ourselves in the case 
$	\ell \ge 0, q\ge 0$ and $\Omega\ge 0$.

%%%%%%%%%%%%%%%%%%%%%%%%%%%%%%%%%%%%%%%%%%%%%
\subsection{Limiting cases}
%%%%%%%%%%%%%%%%%%%%%%%%%%%%%%%%%%%%%%%%%%%%%
Before we obtain general solutions for the 
stationary rotating strings, we see that two limiting 
cases of parameters $\ell, q$, (1) $\ell+q = 1$ and (2) $\ell=0$
(see Figs.\ref{FigHel} and \ref{FigPla}).   

%%%%%%%%%%%%%%%%%%%%%%%%%%%%%%%%%%%%%%%%%%%%%
\subsubsection{Helical strings %($\ell+q = 1$)
}
\begin{figure}[ht]
\begin{center}
\includegraphics[width=12cm]{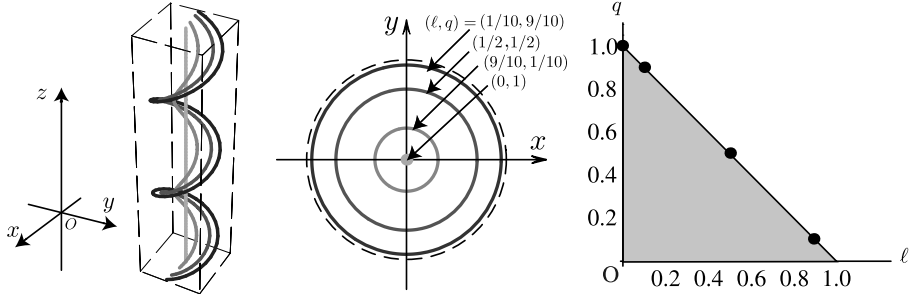}
\end{center}
\caption{Helical strings: 
$ \ell + q =1  (q \neq 0)$. 
The three-dimensional snapshots are given in the left panel, and 
the projection of strings to $x-y$ plane, i.e., projected curves, 
are given in the middle, where $x, y$ are defined by \eqref{cartesian}. 
The dashed circle in the middle figure represents the light cylinder 
$\rho=1/\Omega$. 
In each figure, the strings with the parameters 
$ (\ell, q)=(0,1)$, $(9/10, 1/10)$, $(1/2, 1/2)$, 
$(1/10, 9/10)$ are shown. 
The parameters on $\ell-q$ plane are plotted in the right panel.
}
 \label{FigHel}
\end{figure}

In the case of $\ell+q = 1$, we find from \eqref{minmax} 
that $\rho$ should take a constant value $\rho_0$, where 
\begin{equation}\begin{aligned}
	\rho_0 = \rho_{max}=\rho_{min} 
		=\frac{\sqrt{\ell}}{\Omega}.
\label{amp}
\end{aligned}\end{equation}
When $q\neq 0$, using \eqref{eq-phi} and \eqref{zsigma} 
we see that solutions are described by 
\begin{equation}\begin{aligned}
	&\rho =\rho_0, \quad
  &\varphi =\Omega z.
\label{Helsol} 
\end{aligned}\end{equation}
We call these solutions {\it a helical strings}. The shapes of 
helical strings are shown in Fig.\ref{FigHel}.
In the limit $\ell =0$ i.e., $q=1$, the solutions reduce to 
the straight string solution. 

%%%%%%%%%%%%%%%%%%%%%figure%%%%%%%%%%%%%%%%%
%%%%%%%%%%%%%%%%%%%%%%%%%%%%%%%%%%%%%%%%%%%%%

We can calculate the proper length of snapshot of the string 
with respect to the metric \eqref{eff3metric} as
\begin{equation}\begin{aligned}
		\int ds = \int\sqrt{(1-\Omega^2\rho^2)d\rho^2 + \rho^2 d\varphi^2 
		+(1-\Omega^2\rho^2)d z^2}. 
\end{aligned}\end{equation}
For the helical strings \eqref{Helsol}, the proper length 
for $z$-interval $z \sim  z+\Delta z$ is given by
\begin{equation}\begin{aligned}
\int\sqrt{\rho_0^2 d\varphi^2 
		+(1-\Omega^2\rho_0^2)d z^2} 	
	=\int_{ z}^{ z+\Delta z} d z 
	= \Delta z. 
\end{aligned}\end{equation}
It shows that all helical strings have the same proper length as the straight 
string with the same $ z$-interval. 

In the inertial reference frame \eqref{MinMetCyl}, 
the helical strings are described by
\begin{equation}\begin{aligned}
	&\bar \rho=\rho_0, \\
	&\bar \varphi = \Omega (\bar z+\bar t). 
\label{Helsol2} 
\end{aligned}\end{equation}
The solutions consist of a down-moving wave of 
the angular frequency $\Omega$ and the amplitude $\rho_0$ given by \eqref{amp} 
with the circular polarization. 
The pitch of helix, equivalently the wave length, 
is given by $2\pi \Omega^{-1}$. 
The two-dimensional world surface of helical string 
has another Killing vector 
$\partial_{\bar z}+\Omega\partial_{\bar\varphi}$ which is tangent 
to it in addition to 
$\partial_{\bar t}+\Omega\partial_{\bar\varphi}$. 
It means the world surface of helical string is a homogeneous space.

%%%%%%%%%%%%%%%%%%%%%%%%%%%
\subsubsection{Planar strings %($\ell = 0$)
}
\begin{figure}[ht]
\includegraphics[width=16cm]{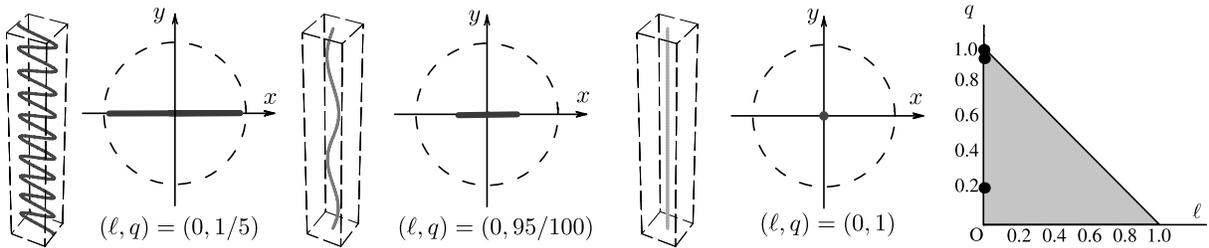}
\caption{Planar strings: $\ell  =0$. 
The three-dimensional snapshots and 
projected curves are shown for $q=1/5$, $95/100$, and $1$ 
as same as Fig.3. 
\\
}
 \label{FigPla}
\end{figure}

%%%%%%%%%%%%%%%%%%%%%%%%%%%
In the case of $\ell = 0$, 
it is convenient to use the rigidly rotating Cartesian coordinate 
\begin{equation}\begin{aligned}
	 x= \rho\cos\varphi, \quad 
 	y= \rho \sin\varphi.
\label{cartesian}
\end{aligned}\end{equation}
From \eqref{eq-phi} we can set $\varphi=0$ ($ y=0$) without loss of 
generality. Then, from \eqref{eq-rho} and \eqref{integral} we have
\begin{equation}\begin{aligned}
	q^2\left(\frac{d x}{d z}\right)^2 = 1-q^2-\Omega^2 x^2. 
\label{eq_planar}
\end{aligned}\end{equation}
The strings are confined in $ x- z$ plane, so 
we call these {\it planer strings}. 
The solutions for \eqref{eq_planar} are 
\begin{equation}\begin{aligned}
 	 x= \rho_{max} \sin \left(\frac{\Omega z}{q} \right),
\label{Plasol}
\end{aligned}\end{equation}
where the amplitude of the waving string $\rho_{max}$ is given by 
the parameters $\Omega$ and $q$ as 
$$
	\rho_{max}=\frac{\sqrt{1-q^2}}{\Omega}. 
$$
The shape of planer strings is depicted in Fig.\ref{FigPla}. 

In the inertial Cartesian coordinate, the planar strings are 
described by
\begin{equation}\begin{aligned}
	\bar x= \rho_{max} \sin \left(\frac{\Omega \bar z}{q}\right)\cos\Omega \bar t, \\
	\bar y= \rho_{max} \sin \left(\frac{\Omega \bar z}{q}\right)\sin\Omega \bar t.
\end{aligned}\end{equation}
These strings consist of a standing wave which is superposition of 
an up-moving wave and a down-moving wave in the equal amplitude. 
In the limit $q=1$ the solution becomes the straight string solution.

%%%%%%%%%%%%%%%%%%%%%%%%%%%
%%%%%%%%%%%%%%%%%%%%%%%%%%%

%%%%%%%%%%%%%%%%%%%%%%%%%%%%%%%%%%%%%%%%%%%%%%%%%%%%%%
\subsubsection{General cases}
%%%%%%%%%%%%%%%%%%%%%%%%%%%%%%%%%%%%%%%%%%%%%%%%%%%%%%

In general, when $\ell\neq 0$, two first order differential equations \eqref{eq-phi} 
and \eqref{eq-rho} are integrated in the form 
\begin{align}
	&{\rho}^2 
		= \frac{1}{2} \left\{
			(\rho_{max}^2+\rho_{min}^2) 
				-(\rho_{max}^2-\rho_{min}^2)\cos\left(2\Omega\sigma\right) 
		\right\} , 
\label{sol_rho} \\
	&\frac{2\ell}{\Omega^2}\tan\left({\varphi} - \varphi_0 
		+ \ell\Omega\sigma\right)
		= 	
	(\rho_{max}^2+\rho_{min}^2)
		\tan\left(\Omega\sigma+\frac{\pi}{4} \right) 
		- (\rho_{max}^2-\rho_{min}^2),
\label{sol_phi} 
\end{align}
where $\varphi_0$ is a constant. (see also ref\cite{BurdenSRS}.)

In the case of $q=0$, \eqref{zsigma} leads that  
strings are confined in a $ z=const.$ plane ($x-y$ plane). 
In this case, the stationarity of the strings breaks down at 
the light cylinder $\rho=\rho_{max}=1/\Omega$. Since the point 
$\rho=1/\Omega$ is curvature singularity, the geodesics have 
end points there. 
This case is a special case of Burden's solution\cite{Burden} 
and discussed in detail by Frolov et al\cite{FSZH}. 
The shape of strings in $x-y$ plane is shown in Fig.5.

%%%%%%%%%%%figure%%%%%%%%%%%%%%%%%
\begin{figure}[ht]
%\centering
\includegraphics[width=8cm]{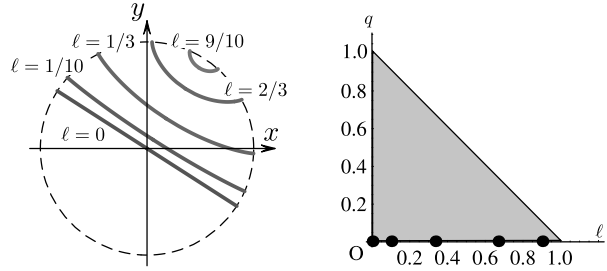}
\caption
{Strings in $x-y$ plane: $q=0$. 
The dashed circle in the left figure is the light cylinder 
$\rho = 1/\Omega$. 
\\
}
 \label{FigHor}
\end{figure}
%%%%%%%%%%%%%%%%%%%%%%%%%%%%%%%%%%%%%%%%%%%%%

Both of the functions $\rho(\sigma)$ and $\varphi(\sigma)$ 
in the solution \eqref{sol_rho} and \eqref{sol_phi} 
are periodic with respect to $\sigma$, but the periods are different. 
It is clear that 
\begin{align}
	\rho(\sigma + \sigma_p ) = \rho (\sigma), 
\end{align}
where $\sigma_p = \pi/\Omega$. 
In this period of $\sigma$, $\varphi$ varies as
\begin{align}
	{\varphi}(\sigma +\sigma_p  ) 
		= {\varphi}(\sigma) + \pi(1-\ell). 
\label{Leg1}
\end{align}
Suppose that $\ell$ is a rational number 
which can be expressed as 
$\ell = a/b$, where $a$ and $b$ are relatively prime. 
During $\sigma$ increases by $n$ times $\sigma_p$, 
$\varphi$ varies as
\begin{align}
{\varphi}(\sigma +n\sigma_p  ) 
		&= {\varphi} (\sigma) + n\pi(1-\ell) \cr
		&= {\varphi} (\sigma) + n\frac{b-a}{b}\pi. 
\label{Leg2}
\end{align}
If the second term in the right hand side holds
\begin{align}
	n\frac{b-a}{b}\pi= 2\pi m, \quad(n,m : \mbox{natural numbers})
\end{align}
$\rho$ and $\varphi$ have a common period. 
The least value of $n$, say $n_\ell$, is determined by 
\begin{align}
	n_\ell &= \frac{2b}{\mathrm{GCD}\bigr[2 b ,(b-a)\bigr] } , 
\label{Nl} 
\end{align}
where $\mathrm{GCD}\left[x,y \right]$ is 
the greatest common divisor of $x,y$.  
The projected curve in the reduced two-dimensional space is closed 
if $\ell$ is rational. The closed curve consists of $n_\ell$ 
elements; each element starts off $\rho_{max}$, evolves through 
$\rho_{min}$, and ends at $\rho_{max}$. 
Till the curve returns to its starting point, 
$\varphi$ laps $2\pi m_\ell$, i.e., the curve 
wraps around the rotation axis $m_\ell$ times, 
where $m_\ell$ is given by
\begin{align}\
 m_\ell = \frac{1-\ell}{2} n_\ell. 
\end{align}
Suppose a closed geodesic in the reduced two-dimensional space 
which consists $n$ elements and which wraps around the rotation axis 
$m$ times. The parameter $\ell$ for this string is given by
\begin{equation}\begin{aligned}
	\ell = 1-\frac{2m}{n}. 
\end{aligned}\end{equation}
When $\ell$ is irrational number, the two-dimensional geodesic 
is not closed. 

The strings with rational $\ell$ have periodic structure 
in $z$ with the period $Z_p=n_\ell q\sigma_p=\pi n_\ell q/\Omega$. On the other hand, 
the strings with irrational $\ell$ have no periodicity. 
We show the shapes of strings 
in Figs. 6 
for $\ell = 1/5$, $1/3$, and $1/2$. 

%%%%%%%%%%%%%%%%%%%%%figure%%%%%%%%%%%%%%%%%
\begin{figure}[ht]
\includegraphics[width=17cm]{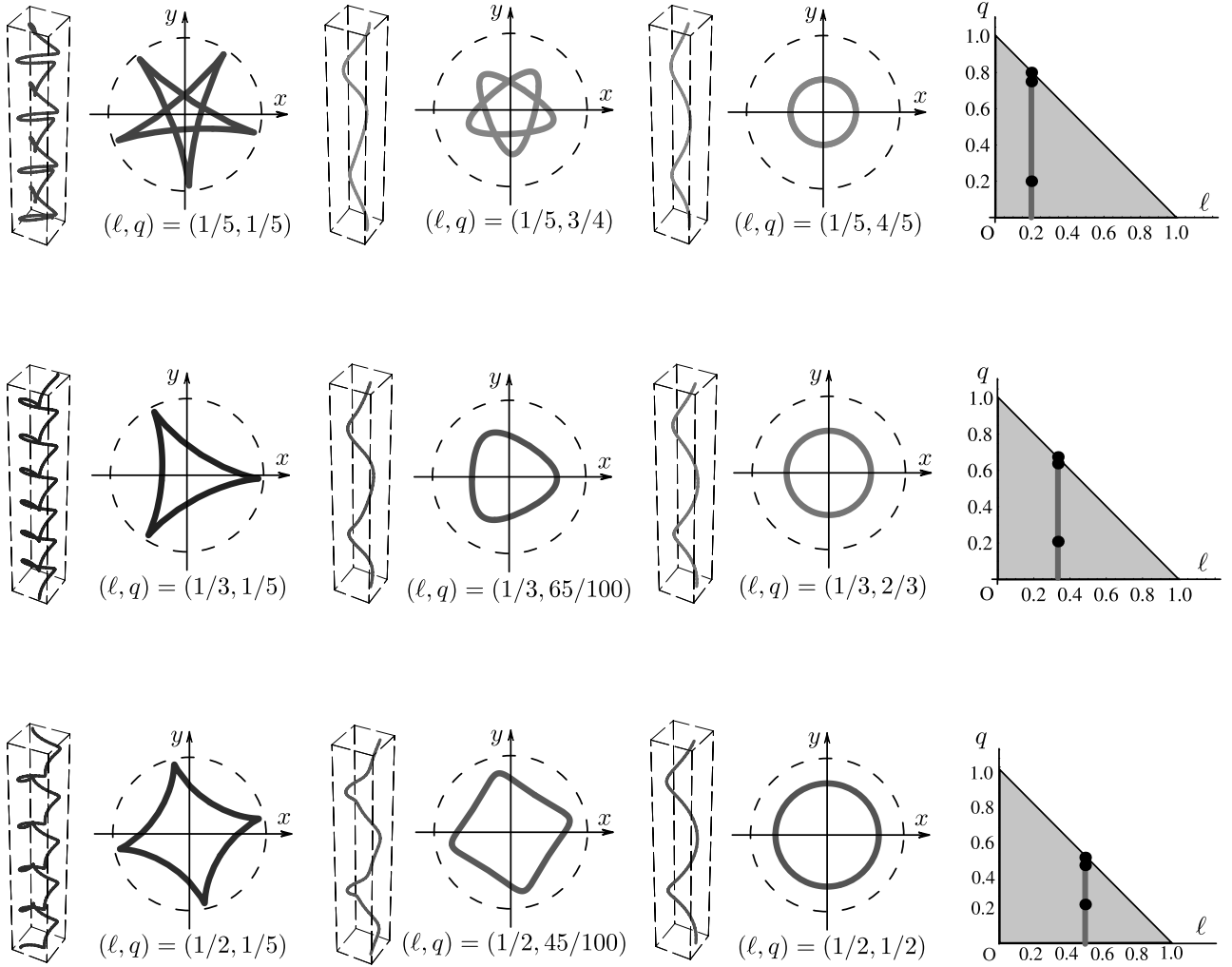}
\caption
{Strings in the cases $\ell = 1/5, 1/3$ and $1/2$. 
\\
}
 \label{Square}
\end{figure}

%%%%%%%%%%%%%%%%%%%%%%%%%
\section{Energy, Momentum and Angular Momentum} 
%%%%%%%%%%%%%%%%%%%%%%%%%

By varying the action \eqref{NG} by $g_{\mu\nu}$ we see that the 
string energy-momentum tensor $T^{\mu \nu}$ is given by 
\begin{align}
	\sqrt{-g} T^{\mu \nu}(x^\lambda)  
		&= -\mu \int d^2\zeta \Theta^{\mu \nu}(\zeta^c) 
		\delta^{(4)}\left(x^\lambda-x^\lambda(\zeta^c) \right), 
\label{EMTofCS0} \\
	&\Theta^{\mu \nu}=\sqrt{-\gamma}\gamma^{ab}\partial_a x^\mu\partial_bx^\nu,
\end{align}
where $x^\lambda(\zeta^c)$ is the solution of $\Sigma$. 

In the inertial reference system \eqref{MinMetCyl}, 
we can rewrite the energy-momentum tenser for the stationary rotating 
strings as 
\begin{align}
	\sqrt{-g} T^{\bar\mu \bar\nu} (\bar t, \bar \rho, \bar \varphi, \bar z)  
	= -\mu \int d\tau d\sigma\ \Theta^{\bar\mu\bar\nu}(\sigma) \
	\delta\left(\bar t-\tau\right)\delta\left(\bar\rho-\bar\rho(\sigma)\right) 
		\delta\left(\bar\varphi-\bar\varphi(\tau, \sigma)\right)
			\delta\left(\bar z-\bar z(\sigma)\right). 
\label{EMTofCS} 
\end{align}
From \eqref{eq-phi}, \eqref{zsigma} and \eqref{integral}, 
we obtain $\Theta^{\bar\mu\bar\nu}$ explicitly. 
(See Appendix B.) 

Now, we define the string energy, $E$, the angular momentum, $J$, 
and the momentum along the rotating axis, $P$. 
We consider infinitely long strings with periodic structure, i.e., $\ell$ 
is assumed to be a rational number, 
then we define $E, J$ and $P$ for one period, $\bar z \sim \bar z+Z_p$ as 
\begin{align}
	E&:= \int_{\rho_{min}}^{\rho_{max}}d\bar\rho\int_0^{2 \pi}d\bar\varphi 
		\int_0^{Z_p} d\bar z 
			\sqrt{-g}~T^{\bar t}_{\ \bar \nu}(-\partial_{\bar t})^{\bar\nu}
		= \mu \int_0^{n_l\sigma_p}d\sigma\ 
			\Theta^{\bar t}_{~{\bar t}}(\sigma), 
\\
	J &:= \int_{\rho_{min}}^{\rho_{max}} d\bar\rho \int_0^{2 \pi} d\bar\varphi \int_0^{Z_p} d\bar z 
				\sqrt{-g}~ T^{\bar t}_{\ \bar \nu}(\partial_{\bar\varphi})^{\bar\nu} 
		= -\mu \int_0^{n_l\sigma_p} d\sigma \ 
			\Theta^{\bar t}_{~{\bar\varphi}}(\sigma), 
\\
	P &:= \int_{\rho_{min}}^{\rho_{max}} d\bar\rho\int_0^{2 \pi} d\bar\varphi 
			\int_0^{Z_p} d\bar z 
			\sqrt{-g}~T^{\bar t}_{\ \bar\nu} (\partial_{\bar z})^{\bar \nu}
		= -\mu\int_0^{n_l\sigma_p}d\sigma\ 
				\Theta^{\bar t}_{~\bar z}(\sigma).
\end{align}
It is easy to calculate these quantities as 
\begin{align}
	E &= \frac{\pi \mu }{|\Omega|} n_\ell (1-\ell^2), 
\label{EDef} \\
	J &= \frac{\pi \mu }{2\Omega |\Omega|} n_\ell (1-\ell^2 -q^2), 
\label{JDef} \\
	P &= - \frac{\pi\mu}{\Omega} n_\ell  \ell q .
\label{PDef}   
\end{align}
We can also define the averaged values of theses quantities per unit length of $z$ as 
\begin{align}
	\braket{E} &:= E/Z_p 
		= \mu \frac{ 1-\ell^2 }{|q|} , 
\\
	\braket{J} &:= J/Z_p 
		=\frac{\mu}{\Omega} \frac{ 1-\ell^2 -q^2 }{2|q|} , 
\label{hEJPDef} \\
	\braket{P} &:= P/Z_p 
		=-\mu \ell \mathrm{sign}(\Omega q).
\end{align}
These quantities are applicable for the strings with irrational $\ell$. 

When $\ell\neq 0$ and $q\neq 0$, we see $P\neq 0 $, that is, 
the strings move along the rotating axis. 
Although the rotating velocity of a string segment is perpendicular 
to the rotation axis, the physical velocity of the Nambu-Goto string 
is orthogonal to the segment. 
Since the strings with $\ell\neq 0$ and $q\neq 0$ have 
inclination to the rotation axis, then the physical velocity of 
string segments have the component along the rotation axis. 
This is the reason for non-vanishing $P$.  
In the planar string case, $\ell=0$, the rotating velocity is physical, 
that is, the velocity is orthogonal to the string because the planar string 
is confined in $ x - z$ plane. Then the planar string does not 
move along the rotation axis. It is consistent with the fact 
that the planar string consists of a standing wave as mentioned before.

In order to see the effective equation of state\cite{VandS} 
for the stationary rotating strings, 
we transform the reference system such that 
the averaged value of momentum $\braket{P}$ vanishes. 
Using the fact that
\begin{equation}\begin{aligned}
	\braket{T^{\bar t\bar t}-T^{\bar z\bar z}} ~\mbox{and}~ 
	\braket{T^{\bar t\bar t}T^{\bar z\bar z}-(T^{\bar t\bar z})^2} 
\end{aligned}\end{equation}
are invariant under
the Lorentz transformation along the $\bar z$-axis, 
we obtain effective line density, $\tilde\mu$, 
and effective tension, $\tilde {\cal T} $, as
\begin{equation}\begin{aligned}
	&\tilde\mu %= <T^{tt}> 
		=\frac{\mu}{2|q|}
				\left[{1-\ell^2+q^2}
				+\sqrt{(1-q-\ell)(1-q+\ell)(1+q-\ell)(1+q+\ell)}\right],\\
  &\tilde {\cal T} %= <T^{zz}> 
		=\frac{\mu}{2|q|}
				\left[{1-\ell^2+q^2}
				-\sqrt{(1-q-\ell)(1-q+\ell)(1+q-\ell)(1+q+\ell)}\right].
\end{aligned}\end{equation}
In general, it holds that $\tilde \mu\tilde {\cal T}= \mu^2$
and $\tilde \mu \geq \tilde {\cal T}$. 
In the case of helical strings, there exists no inertial reference system 
such that $\braket{P}$ vanishes because a single wave moves with the velocity 
of light along the rotating axis. 

%%%%%%%%%%%%%%%%%%%%%%%%%%%%%%%%%%%%%%%%%%%%%%%%%%%%%%%%%%%%%%%%%
\section{Conclusion}
%%%%%%%%%%%%%%%%%%%%%%%%%%%%%%%%%%%%%%%%%%%%%%%%%%%%%%%%%%%%%%%%%

We study stationary rotating Nambu-Goto strings in Minkowski spacetime. 
It has been shown that the stationary rotating strings with an angular 
velocity $\Omega$ are described by geodesics in two-dimensional curved 
spaces with positive definite metrics with 
a parameter, $q$ in this paper. 
The metrics admit a Killing vector which generates rotation symmetry, 
then the geodesics in the two-dimensional spaces have a constant of motion, 
$\ell$ in this paper. 
Therefore, the stationary rotating strings in Minkowski spacetime 
are characterized by three parameters $(\Omega, q, \ell)$. 

One of the typical stationary rotating strings are the 
\lq {\it helical strings}\rq , 
$|q|+|\ell|=1$, where a snapshot looks a helix. The world surface of the helical 
string is the two-dimensional homogeneous space embedded in Minkowski spacetime. 
For general $(\Omega,q,\ell)$, stationary rotating strings have 
quasi-periodic structure along the rotation axis. 
Only in the case of rational $\ell$, the strings have 
exact periodicity. The strings display much variation 
in the shape which depends on $q$ and $\ell$. 

We have obtained the following averaged values per unit length 
along the rotation axis: 
energy, angular momentum and linear momentum along the axis. 
It should be noted that 
the rotation of the strings around the axis induces the linear momentum 
along the axis because of the inclination of string segments. 

By the advantage of the analytic solutions of the string motion, 
we can calculate the energy-momentum tensor easily. 
Then, we can study the gravitational field yielded by the stationary 
rotating strings, especially gravitational wave emission. 
It is important to clarify the property of the gravitational waves from the 
stationary rotating strings for verification of their existence 
in the universe. We will report this issue in near future\cite{WGE2}. 

It is also interesting problem 
to construct all cohomogeneity-one string 
in Minkowski\cite{IshiharaKozaki, KKImink}, de Sitter, 
and anti-de Sitter\cite{KKIads} spacetimes.
Furthermore, it would be challenging work 
to find the general solutions of stationary rotating strings 
in the black hole spacetimes as an extended work of ref.\cite{FSZH}.

%%%%%%%%%%%%%%%%%%%%%%%%%%%%%%%%%%%%%%%%%%%%%%%%%%%%%%%%%%%%%%%%%%%%%%%%%%%%%%%
\section*{Acknowledgments}
%%%%%%%%%%%%%%%%%%%%%%%%%%%%%%%%%%%%%%%%%%%%%%%%%%%%%%%%%%%%%%%%%%%%%%%%%%%%%%%%
%%%
We would like to thank K. Nakao and C.-M. Yoo for useful discussions. 
HN is supported by a JSPS Research Fellowship 
for Young Scientists, No.~5919.
He is also supported by JSPS for Research Abroad 
and in part by the NSF through grants PHY-0722315, 
PHY-0701566, PHY-0714388, and PHY-0722703, 
and from grant NASA 07-ATFP07-0158.

%%%%%%%%%%%%%%%%%%%%%%%%%%%%%%%%%%%%%%%%%%%%%%%%%%%%%%%%%%%%%%%%%%%%%%%%%%%%%%%
\appendix
%%%%%%%%%%%%%%%%%%%%%%%%%%%%%%%%%%%%%%%%%%%%%%%%%%%%%%%%%%%%%%%%%%%%%%%%%%%%%%%%
\section{Potential in Two-dimensions}

We consider a particle in the two-dimensional flat space driven 
by a potential force. 
In general, every metric of two-dimensional space is written in conformal flat form. 
The metric \eqref{2dim-metric} is written in the form
\begin{equation}\begin{aligned}
  ds_{red}^2 
	&= (1-q^2-\Omega^2\rho^2)
	\left(d\rho^2 +\frac{\rho^2}{1-\Omega^2\rho^2}d\tilde\varphi^2\right)\\
	&= F(x^A) \left(d r^2 +r^2 d\varphi^2\right),
\label{conformal_metric}
\end{aligned}\end{equation}
where
\begin{equation}\begin{aligned}
  &F= \frac{1-q^2-\Omega^2\rho^2}{1-\Omega^2\rho^2}
	(1+\sqrt{1-\Omega^2\rho^2})^2 e^{-2\sqrt{1-\Omega^2\rho^2}}, 
\label{conformal}\\
	&r=\frac{\rho}{1+\sqrt{1-\Omega^2\rho^2}} e^{\sqrt{1-\Omega^2\rho^2}}. 
\end{aligned}\end{equation}

\begin{figure}[ht]
 \includegraphics[width=12cm ,clip]{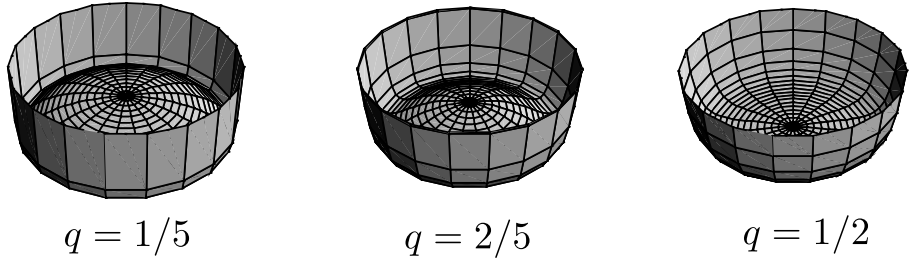}
\caption{Potentials for $q=1/5, 2/5$ and $1/2$ in two-dimensional 
flat space.
}
\label{potential}
\end{figure}

Using the argument from \eqref{Sred} to \eqref{hred} inversely, 
we see that geodesics in the metric \eqref{conformal_metric} is equivalent 
to the orbits of the particle whose dynamics is governed by the action
\begin{equation}\begin{aligned}
	S = \int \left[p_A {x}'^A 
			- N_{(2)}{\cal H} \right] d\sigma 
\end{aligned}\end{equation}
with
\begin{equation}\begin{aligned}
	&{\cal H} = \frac12\delta^{AB}p_A p_B + U(r), 
\\
	&U(r)=-\frac12 F(\rho(r)), 
\end{aligned}\end{equation}
where the particle should satisfy the Hamiltonian constraint ${\cal H} =0$. 
The shape of potential $U(r)$ is shown in Fig.\ref{potential}.

%%%%%%%%%%%%%%%%%%%%%%%%%%%%%%%%%%%%%%%%%%%%%%%%%%%%%%%%%%%%
%%%%%%%%%%%%%%%%%%%%%%%%%%%%%%%%%%%%%%%%%%%%%%%%%%%%%%%%%%%%
\section{The components of $\Theta^{\mu\nu}$}

The components of $\Theta^{\mu\nu}$ are explicitly expressed in the following:
\begin{equation}\begin{aligned}
\Theta^{tt}  &= -(1-\ell^2), \\
\Theta^{t\rho}  &= \frac{\ell\, \Omega }{2 \rho }
(\rho_\mathrm{max}^2 -\rho_\mathrm{min}^2) \sin \left(2|\Omega|\sigma \right),  \\
\rho  \, \Theta^{t\varphi}  
	&= -\frac{\Omega^2 \rho^2  - \ell^2}{\Omega \rho }, \\
\Theta^{tz}  &= q\ell \mathrm{sign}(\Omega), \\
\Theta^{\rho\rho}  
	&= \frac{\Omega^2 }{4\rho^2 } 
 (\rho_\mathrm{max}^2 -\rho_\mathrm{min}^2)^2\sin^2 \left(2|\Omega|\sigma\right), \\
\rho  \, \Theta^{\rho\varphi}  
	&= \frac{\ell \, }{2\rho^2}
 (\rho_\mathrm{max}^2 -\rho_\mathrm{min}^2)\sin \left(2|\Omega|\sigma\right),  \\
\Theta^{\rho z}  
	&= \frac{q|\Omega|}{2\rho}(\rho_\mathrm{max}^2 -\rho_\mathrm{min}^2)
 		\sin \left(2|\Omega|\sigma\right),  \\
\rho^2  \,\Theta^{\varphi \varphi}  
	&= -\Omega^2 \rho^2
\left( 1- \frac{\ell^2}{\Omega^4 \rho^4 } \right), \\
\rho  \, \Theta^{\varphi z}  
	&= \frac{\ell q}{|\Omega| \rho },  \\
\Theta^{zz}  &= q^2. 
\label{Thetamunu} 
\end{aligned}\end{equation}

In these expressions, $\rho=\rho(\sigma)$ is given by \eqref{sol_rho}.

\newpage
%\begin{references}
%%%

%\end{references}

%%%%%%%%%%%%%%%%%%%%%%%%%%%%%%%%%%%%%%%%%%%%%%%%%%%%%%%%%%%%%%%%%%%%%%%%%%%%%%%%

\end{document}